\documentclass[aps,twocolumn,amsfonts,prl,nofootinbib,showpacs]{revtex4}
\usepackage{graphicx}
\usepackage{amsmath}

\usepackage{color}

\def\bra{{\langle}}
\def\ket{{\rangle}}

\def\b{\beta}
\def\Om{\Omega}
\def\la{\lambda}
\def\be{\begin{equation}}
\def\ee{\end{equation}}
\newcommand{\eea}{\end{eqnarray}}
\newcommand{\bea}{\begin{eqnarray}}

\def\g{\gamma}
\def\d{\delta}

\def\om{\omega}

\def\dw{\dot w }

\usepackage{bm}
\usepackage{epsfig}

\begin{document}

\title{Running-mode resonance in A.C.-biased periodic potential}

\author{Lance Labun}
\email{labun@physics.arizona.edu}
\altaffiliation{Now at: Department of Physics, University of Arizona, Tucson, AZ 85721, USA}
\affiliation{Department of Physics and Astronomy,
Dartmouth College, Hanover, NH 03755, USA}

\author{Marcelo Gleiser}
\email{marcelo.gleiser@dartmouth.edu}
\affiliation{Department of Physics and Astronomy, 
Dartmouth College, Hanover, NH 03755, USA}

\begin{abstract}
We investigate the stochastic dynamics of a particle in the presence of a modulated sinusoidal potential. Using the time derivative of the winding number, we quantify the particle's motion according to its running time, the time it runs monotonically to the left or right. For a range of model parameters, we show that, in the overdamped regime, the particle's motion in this modulated washboard potential exhibits stochastic resonance. We briefly suggest possible applications of our results, including the amplification of signals for measurement devices and in stimulated tunneling of Bose-Einstein condensates.
\end{abstract}
 
\maketitle

\begin{section}{Introduction}
Stochastic resonance (SR), originally proposed in reference to two equally-favored climate states, has only more recently been discussed as a more general concept independent of an explicitly bistable potential function (see~\cite{SRRev} for a review).  It has now been identified in a variety of different contexts, from lasers and optical traps to neurological systems and biochemical reactions.  Quantum versions of stochastic resonance have also received attention, especially in superconducting quantum interference devices (SQUIDs) and nanomechanical resonators due to their location near the boundary of classical and quantum dynamics.  It has been shown that SR can be an efficient amplification mechanism, and similar ideas are of interest to controlled diffusion, for instance by applying various noise models in symmetric or asymmetric periodic potentials to obtain directed transport (such as described by ``Brownian motors"~\cite{BMotors}).  

The interest in nonlinear resonance phenomena in periodic potentials is not due solely to models of diffusion and transport, as such potentials underlie other interesting physical systems, for example, sphalerons in high energy physics, which have been invoked in the context of the electroweak phase transition \cite{sphalerons}.  Sticking to the concrete language of diffusion, many authors have discussed SR in the basic sinusoidal potential, starting with the simplest (static, symmetric) case ~\cite{Melnikov, JumpLength}, implementing a static bias~\cite{LongJumps, Comment, ForwardBack}, and finally tackling actively modulated systems~\cite{QDiff, ExpActDyn, NJP, NoiseMix}.  Borromeo and Marchesoni recently generalized stochastic resonance to dynamically-defined states in a tilted washboard potential~\cite{ACdrive}.

One case not yet treated in the literature is a rocking washboard, in which a periodic potential is subjected to a global modulation, producing alternately positive and negative biases.  This system offers a simple model for diffusion in the presence of agitation, such as experienced by adatoms or particulates on a vibrated surface, or even electrons on the surface of a dielectric and subject to a radiation field.  Periodic potentials are also known in particle physics, so the possibility of an oscillatory bias, e.g., through a background scalar field \cite{srivastava}, further recommends the relevance of this model.  In the next section we introduce the model and explore some of its properties. In section 3 we discuss our results, while in section 4 we present a summary.

\end{section}

\begin{section}{Model}

The equation of motion is the usual one associated with studies of Brownian motion and stochastic resonance,
\be \label{geneom}
\ddot x =  \xi(t) -\g \dot x -V'(x,t)
\ee
in rescaled units.  The first two terms on the right represent the thermal bath at temperature $T$ via the fluctuation-dissipation theorem: Gaussian white noise of zero mean $\bra \xi(t) \ket = 0$ and $\d$ correlation function $\bra \xi(t)\xi(t') \ket = 2\g kT \d(t-t')$ combined with viscous damping $\g$.  $k$ is Boltzmann's constant. The last term is the potential, which has both a static and a periodic driving component
\be \label{pot}
V(x,t)=-(A/2\pi )\cos (2\pi x) + xF\cos \Om t.
\ee
Previous work \cite{QDiff, ExpActDyn, NJP} has maintained a constant bias throughout any time-dependent modulations of barrier heights, so that the driving component appeared as $F(t) = F_0 + F_1 \cos \Om t$ for $F_0 > 0$.  Our study departs by setting $F_0 =0$, thereby tilting the whole washboard rather than merely modulating its corrugation.

We consider a low frequency input and the overdamped limit, in which the equation of motion (\ref{geneom}) can be adiabatically reduced to 
\be \label{odeom}
\dot x = -A \sin (2\pi x) + F\cos \Om t + \xi(t)
\ee
with $t$ normalized to absorb $\g$.  The potential was defined so that integer values of the position correspond to the winding number $\omega$.
 
In the limit of slow modulation, $\Om \ll 1$, transitions between wells are thermally induced for any value of the damping.  At various points in its evolution, the potential has a nearly constant bias in the positive or negative direction, and the difference induced in the right and left barrier heights is given by
\be \label{DVt}
V(1/2,t)-V(-1/2,t) = F \cos\Om t.
\ee
For phases near $\Om = \pi/2, 3\pi/2$, regular diffusion across an unbiased periodic potential applies, while around $\Om = 0, \pi$, the dynamics are effectively those of activated diffusion. In the latter case, the extensive characterization of the jump length and time statistics given by \cite{JumpLength, UnbiasedSin} for the tilted washboard potential may be straightforwardly applied.  Jump lengths and times, respectively defined as the number of wells and the total period of monodirectional motion, are controlled by the ratio $F/\g$, as discussed in~\cite{ACdrive}.  
 
The activated behavior near maximum bias will only dominate the overall statistics of the system if the particle is effectively locked into a specific well for the remainder of the phase.  Letting $E^- \equiv A/\pi -F/2$, the reduced barrier height in the direction of the bias, left at $\Om = 0$ and right at $\Om = \pi$, we may derive an intuitive constraint in the spatially-diffusion-controlled regime (large damping in our normalization) via Kramers rates \cite{RateTheory}.  On the one hand, the particle should experience the tilted washboard potential for longer than the probable escape period, $T_{min} < T_{\Om}/4$ with $T_{min} = 1/r_k = (2\pi\g/\om_0\om_b)\exp (\b E^-)$ the inverse of the Kramers rate near maximum bias and the quarter period being a reasonable choice.  ($\om_0$ and $\om_b$ correspond to the oscillating frequencies at the metastable well and top of the barrier, respectively, and $\b=1/kT$.) On the other hand, the particle should also have a suppressed probability of escape near zero bias, and hence $T_{max} > T_{\Om}/4$ for $T_{max} = (2\pi\g/\om_0\om_b)\exp(\b A/\pi )$.  Rearranging and combining the inequalities, we find
\be \label{odineq}
0 < (kT/A) \ln (4\alpha_o) + 1/\pi < F/2A
\ee
where $\alpha_o\equiv \Om\g/\om_0^2$ is a dimensionless time parameter in which we used the fact that $\om_b = \om_0$ for the sinusoidal potential.  Because this inequality locates the region of parameter space in which we expect the activated behavior to dominate the overall dynamics, we clearly desire $F/A$ as large as possible.

A very similar constraint, more accurate in the case of low damping, can be derived from the Kramers rate in the energy-diffusion-limited regime, $\g I(E) \ll kT$ (in which $I(E)$ is the energy-dependent flux in action space).  An extra step is required in combining the inequalities due to the different values $I(E^-)$ and $I(A/\pi )$,
\begin{subequations}
\begin{eqnarray} \label{odstep}
I(A/\pi) \frac{\om_0}{4\g\Om}e^{\b A/\pi } & < & 1, \\
I(E^-) \frac{\om_0}{4\g\Om}e^{\b A/\pi } & > & e^{\b F/2}.
\end{eqnarray}
\end{subequations}
We then use the monotonicity of $I(E)$, from which $I(E^-) > I(A/\pi)$, to write
\be 
0 < (kT/A) \ln  (4 \alpha_u)  + 1/\pi < F/2A.
\ee
We retain the larger flux $I(E^-)$ in order to strengthen former inequality $(\ref{odstep})$, since it expresses the suppression of the flux during the unbiased parts of the period.
For the effect to become pronounced, the inequality should clearly be as strong as possible.  However, in the context of stochastic resonance, small input biases are of greater interest, and in the following $F<A$.
 
In a biased periodic potential, overdamped systems represented by (\ref{odeom}) can exhibit a consistently running state only for $F=2A$, as the barriers are completely suppressed in this case.  However, many results for diffusion in a tilted periodic potential are quite general~\cite{ODMobility}, and directed diffusion still arises with an average velocity given by
\be
\bra \dot x \ket = \frac{(kT/\g)(1-\exp(-F/kT)}{\int\int^1_0 dx dy \exp[(V(x)-V(x-y)-yF)/kT]}.
\ee
Considering shorter time dynamics, the modulation of the barrier heights introduces a phase bias for diffusive transitions.  For small $F$, a simple correlation should be apparent in the residence time distribution, as jumps between wells will tend to occur in phase with the modulation.

We therefore expect a strong correlation between the direction of transport and the phase of the forcing.  Given that transitions are suppressed when the biasing force $F(t)=F\cos\Om t$ is near zero, transitions against the proffered bias are even more strongly suppressed (though velocity inversions occur less often than initially expected in a statically biased potential according to~\cite{ForwardBack}).
 
We observed a dynamic resonance between the time derivative of the winding number $\dot w$ and the modulation $F(t)$, now understood as an input signal.  Defining  ``left-running'' and ``right-running'' states corresponding to $\dw < 0$ and $\dw > 0$, respectively, the formalism and language of stochastic resonance can be applied.  All of the generally accepted ingredients are present: (i) an activation barrier, $E^- \equiv 2A-F/2$; (ii) a (weak) input signal, the global modulation of the potential, $F(t)$ in~(\ref{DVt}); and (iii) a source of noise, namely the stochastic term in~(\ref{geneom}).
\end{section}

\begin{section}{Results}
The most direct connection to previous observations of SR lies in residence times, defined as the time spent in a given well:  we found that the distribution of residence times for the left- and right-running states---or, more appropriately to our situation, the {\it running times}---are peaked around the half integer multiples of the forcing period, $(n+\frac{1}{2})T_{\Om}$.  Running times for various values of the temperature are plotted in figure~\ref{ODtimes}, and the qualitative similarity to plots for resident times such as found in~\cite{SRRev} is easily verified.

\begin{figure}
\includegraphics{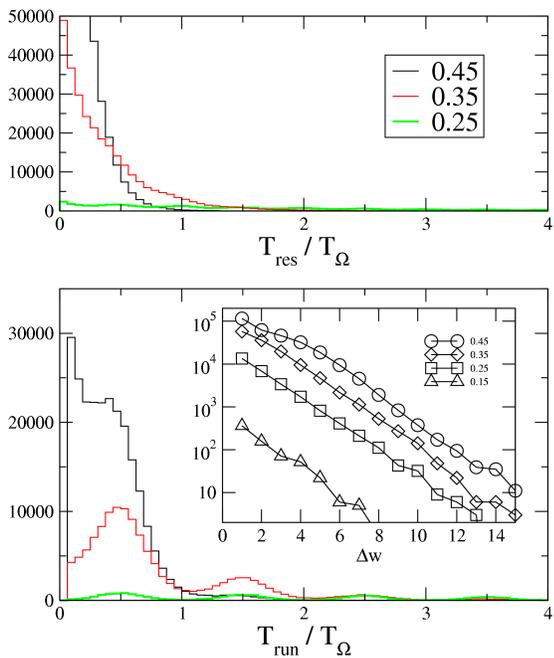}
\caption{ \label{ODtimes} The residence (top) and running (bottom) time distributions in the overdamped model for three temperatures $2\pi kT/A=0.25, 0.35, 0.45$.  The vertical axis denotes  the number of instances of a measured period, normalized on the horizontal axis by the forcing period, $T_{\Om}$.  The residence time also displays a (barely-discernable) resonance, showing a preference for transitions on both integer and half-integer multiples of the period at sufficiently low temperature.  The distribution of the jump length $\Delta w$ in the inset shows the expected exponential die off~\cite{JumpLength}.  Parameters: $A=10^4, F=2000, \Om=10$}
\end{figure}

We note that the usual resonance in the {\it residence time} distribution (time spent localized in a given well) can still be identified at low temperatures, although not as clearly due to the scale of the figure. Below, we will present results in the underdamped regime, where the resident time resonance is more apparent (see Fig. 4).

With $\g$ normalized to 1, $F/\g$ is always less than $8\sqrt{A}$ (the threshold for the locked regime~\cite{ACdrive}), so jump lengths display the expected exponential distribution, as can be seen in the inset of figure~\ref{ODtimes}~\cite{JumpLength}. 

Note that `running' refers only to the monotonicity of the winding, as the particle almost always resides a time comparable to the relaxation time $\om_0^{-1}$ even when only transiting a well.  For this reason, in order to obtain the clean results displayed in figure~\ref{ODtimes}, we considered a change in winding number to occur only after the particle begins its descent into the adjacent well, not merely when it has surmounted the barrier.  This definition is the cause of the low frequency cutoff apparent upon close inspection.

Further support for characterizing the running time as a resonance comes from the Fourier transform of $x$.  The signal-to-noise ratio should show a maximum in the temperature range suggested by~(\ref{odineq}).  Figure~\ref{FTOD} displays the Fourier transform of a system in the suggested range and furthermore exhibits the first harmonic at $3\Om$ as predicted by~\cite{FTHarmonics}.

\begin{figure}
\includegraphics{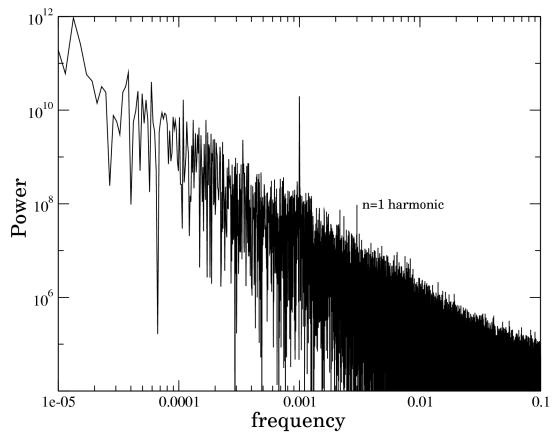}
\caption{ \label{FTOD}  The Fourier transform of position $x$ in an overdamped system, driven at $\Om=10^{-3}$.  The primary peak at the driving frequency is clear, accompanied by an extra peak at the first harmonic, $3\Om$, exactly as predicted by~\cite{FTHarmonics}.  Parameters: $A=1, F=.3, 2\pi kT=.325$}
\end{figure}

In~\cite{NoiseMix}, it is noted that diffusion is somewhat suppressed in a tilted periodic potential (as compared to Einstein diffusion $D_0=kT/\g$), and the global modulation may be expected to suppress diffusion even further by introducing an extra coherence in the transport of particles in the potential.  As can be seen in figure~\ref{Diff} we have investigated this fact, finding a very significant suppression, which decreases with increasing temperature.

\begin{figure}
\includegraphics{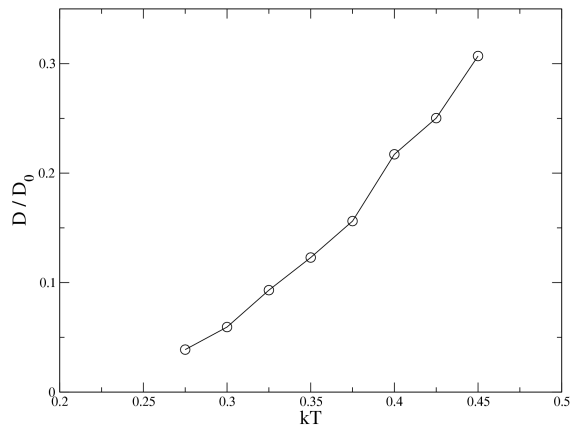}
\caption{ \label{Diff}  The normalized diffusion coefficient $D/D_0$ plotted against temperature, obtained from the regression of $\bra x^2 \ket$ from 1000 particles.  Parameters: $A=1, F=.3, T_{\Om}=1000$}
\end{figure}

In underdamped systems $\g \ll \om_0$, the left- and right-running states have an obvious counterpart in the jump times and lengths discussed in~\cite{ACdrive}.  However, the correlation of $\dw$ to the phase of the drive cannot be properly called a resonance phenomenon, as inertial effects reinforce rather than suppress the peaked distribution, as demonstrated in figure~\ref{uddist}.  Physically, if $\g$ is of the same order as $2\Om$, sufficient kinetic energy may be conserved through the half period of opposing bias such that no velocity inversion occurs~\cite{ForwardBack}.  Evidence for this explanation is derived from the Fourier transform of $x$: the power in peaks at higher harmonics should generally {\it increase} with temperature, as lower frequency velocity inversions become more common.  Simulations support this, to a point, as we find that the peaks actually become saturated beyond a certain temperature, as shown in figure~\ref{powerud}.  Although the saturation indicates a decrease in signal-to-noise ratio, it does not produce a profile sufficiently peaked as to justify its characterization as a resonance.

\begin{figure}
\includegraphics{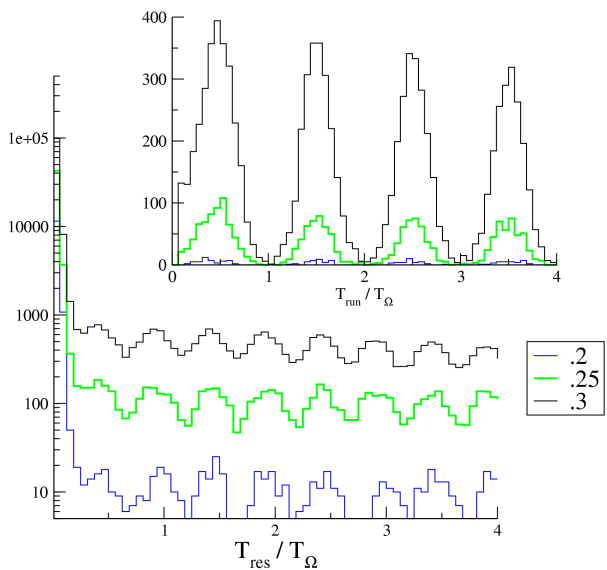}
\caption{ \label{uddist} Residence (bottom) and running (top) time distributions for an underdamped system at temperatures $2\pi kT=0.2, 0.25, 0.3$.  The response of $\dw$ does {\it not} show a resonance-like behavior, as the distribution remains sharply peaked at the half integer multiples even while temperature increases.  The residence time is also dramatically peaked near zero, though the higher harmonics here also gain power as temperature increases.  The steadiness of the distribution, however, highlights that it is entirely an inertial effect.  Parameters: $A=1, F=0.3, T_{\Om}=100$ }
\end{figure}

\begin{figure}
\includegraphics{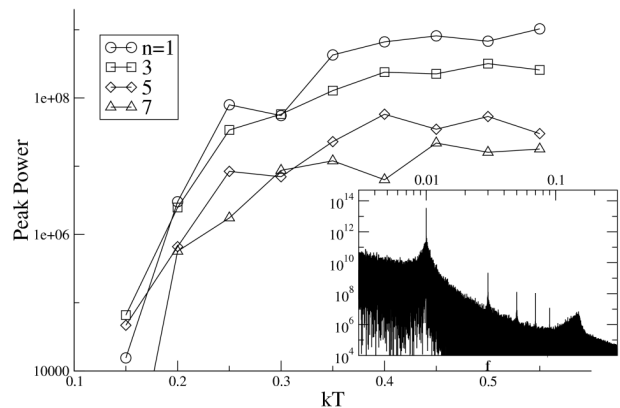}
\caption{ \label{powerud}  The power in the delta peaks of the first through fourth harmonics, $(2n+1)\Om$ for $n=1..4$.  The power in each peak increases nonlinearly with temperature, indicating an increase in signal to noise ratio---behavior contrary to the resonant character of SR.  Each harmonic appears to saturate around $2\pi kT=0.4$.  The inset displays the Fourier transform of an experimental run at $2\pi kT=0.35$ in which the peaks at the higher harmonics can be clearly identified.  Parameters: $A=1, F=0.3, \g=0.015, T_{\Om}=100$  }
\end{figure}

\end{section}

\begin{section}{Summary}
We described a dynamic resonance exhibited by an overdamped particle in a rocking washboard potential. Defining ``running'' states according to the sign of $\dw$, whereby a particle runs monotonically to the left or right before reversing its direction, we were able to characterize the phenomenon as a new kind of stochastic resonance. Numerical results are consistent with the range of parameters derived from naive expectations. We have also shown that no such resonances are possible in the underdamped regime.
 
Looking ahead, our results are of interest to measurement considerations in the context of bifurcation amplification~\cite{BifAmp}, whereby a small input signal is enhanced so as to effect the  distinction between the (bistable) dynamical states arising just beyond the bifurcation threshold of a nonlinear system.  This immediately suggests a possible direction in which these ideas may be taken: instead of $F(t)$ being an unvarying input mode, the particle may be coupled to a second system whose oscillations provide the input signal.  From this perspective, $(\ref{geneom})$ represents the high mass-ratio limit of this second system, so that the observed variable (modeled here) effects no significant back-reaction on the input signal.  For example, starting from the equations of motion for a bilinearly coupled pair
\be \begin{split}
m_1\ddot x_1^2 & = -V_1'(x_1)-\g_1\dot x_1 + \la x_2 \\
m_2\ddot x_2^2 & = -V_2'(x_2)-\g_2\dot x_2 + \la x_1
\end{split} \ee 
we may consider the perturbative case in which the solution $x_2(t)$ is harmonic, being limited to a region near one of its potential minima.  Secular perturbation theory on the full system produces the expected condition for the validity of this Ansatz, namely $\la / m_2 \ll 1$.  Then with $(\ref{geneom})$ a model of such a bipartite system, the dynamical resonance described above becomes a large amplitude signal for the oscillation of the underlying system ($x_2$) between two of its states.  These states may be dynamical or static, and the coupling engineered accordingly.

A second connection can be made to recent advances in quantum tunneling dynamics.  Experiments were recently undertaken in which resonantly enhanced tunneling was clearly observed in a Bose-Einstein condensate in a tilted washboard potential~\cite{RET}.  Although the proper quantum mechanical calculations should be carried out in detail, the classical results here suggest that by tuning the oscillation of the washboard to the tunneling rate, a second order resonance should arise.  The loss of particles to the continuum following the first tunneling event could be drastically reduced by reversing the tilt on the potential.  The discretization of the energy levels provides an inherent suppression of tunneling events during the inversion of the bias.  Testing this hypothesis might more easily be achieved in a circularly symmetric optical potential such as was used in~\cite{CircBrownMotion}, so that the modulated bias is alternately positively and negatively radial.

This paper demonstrates the utility of stochastic resonance as a general model of phenomena for which obvious localized states are not necessarily presented by the form of the potential.  Clearly, with ideas progressing in such directions as bifurcation amplification~\cite{BifAmp} and the ability to easily construct and manipulate periodic potentials thanks to laser trapping, the language and the theory to describe the possible dynamics should develop as well.  Stochastic resonance in the context of more generalized states will almost certainly appear and be fruitfully discussed in ways similar to those presented above.
\end{section}

\quad

We thank Miles Blencowe for his thoughts and support.  This work was supported by a summer research grant from the Department of Physics and Astronomy at Dartmouth College. MG is supported in part by a National Science Foundation grant PHY-0653341.

\bibliography{draft}

\begin{thebibliography}{21}
\expandafter\ifx\csname natexlab\endcsname\relax\def\natexlab#1{#1}\fi
\expandafter\ifx\csname bibnamefont\endcsname\relax
  \def\bibnamefont#1{#1}\fi
\expandafter\ifx\csname bibfnamefont\endcsname\relax
  \def\bibfnamefont#1{#1}\fi
\expandafter\ifx\csname citenamefont\endcsname\relax
  \def\citenamefont#1{#1}\fi
\expandafter\ifx\csname url\endcsname\relax
  \def\url#1{\texttt{#1}}\fi
\expandafter\ifx\csname urlprefix\endcsname\relax\def\urlprefix{URL }\fi
\providecommand{\bibinfo}[2]{#2}
\providecommand{\eprint}[2][]{\url{#2}}

\bibitem[{\citenamefont{Gammaitoni et~al.}(1998)\citenamefont{Gammaitoni,
  H\"{a}nggi, Jung, and Marchesoni}}]{SRRev}
\bibinfo{author}{\bibfnamefont{L.}~\bibnamefont{Gammaitoni}},
  \bibinfo{author}{\bibfnamefont{P.}~\bibnamefont{H\"{a}nggi}},
  \bibinfo{author}{\bibfnamefont{P.}~\bibnamefont{Jung}}, \bibnamefont{and}
  \bibinfo{author}{\bibfnamefont{F.}~\bibnamefont{Marchesoni}},
  \bibinfo{journal}{Rev. Mod. Phys.} \textbf{\bibinfo{volume}{70}},
  \bibinfo{pages}{223} (\bibinfo{year}{1998}).

\bibitem[{\citenamefont{H\"{a}nggi et~al.}(2005)\citenamefont{H\"{a}nggi,
  Marchesoni, and Nori}}]{BMotors}
\bibinfo{author}{\bibfnamefont{P.}~\bibnamefont{H\"{a}nggi}},
  \bibinfo{author}{\bibfnamefont{F.}~\bibnamefont{Marchesoni}},
  \bibnamefont{and} \bibinfo{author}{\bibfnamefont{F.}~\bibnamefont{Nori}},
  \bibinfo{journal}{Ann. Phys.} \textbf{\bibinfo{volume}{14}},
  \bibinfo{pages}{51} (\bibinfo{year}{2005}).

\bibitem[{\citenamefont{Arnold and McLerran}(1988)}]{sphalerons}
\bibinfo{author}{\bibfnamefont{P.}~\bibnamefont{Arnold}} \bibnamefont{and}
  \bibinfo{author}{\bibfnamefont{L.}~\bibnamefont{McLerran}},
  \bibinfo{journal}{Phys. Rev. D} \textbf{\bibinfo{volume}{37}},
  \bibinfo{pages}{1020} (\bibinfo{year}{1988}).

\bibitem[{\citenamefont{Melnikov}(1991)}]{Melnikov}
\bibinfo{author}{\bibfnamefont{V.}~\bibnamefont{Melnikov}},
  \bibinfo{journal}{Phys. Rep.} \textbf{\bibinfo{volume}{209}},
  \bibinfo{pages}{1} (\bibinfo{year}{1991}).

\bibitem[{\citenamefont{Ferrando et~al.}(1993)\citenamefont{Ferrando,
  Spadacini, and Tommei}}]{JumpLength}
\bibinfo{author}{\bibfnamefont{R.}~\bibnamefont{Ferrando}},
  \bibinfo{author}{\bibfnamefont{R.}~\bibnamefont{Spadacini}},
  \bibnamefont{and} \bibinfo{author}{\bibfnamefont{G.}~\bibnamefont{Tommei}},
  \bibinfo{journal}{Phys. Rev. E} \textbf{\bibinfo{volume}{48}},
  \bibinfo{pages}{2437} (\bibinfo{year}{1993}).

\bibitem[{\citenamefont{Reimann et~al.}(2002)\citenamefont{Reimann, den Broeck,
  Linke, H\"{a}nggi, Rubi, and P\'{e}rez-Madrid}}]{LongJumps}
\bibinfo{author}{\bibfnamefont{P.}~\bibnamefont{Reimann}},
  \bibinfo{author}{\bibfnamefont{C.~V.} \bibnamefont{den Broeck}},
  \bibinfo{author}{\bibfnamefont{H.}~\bibnamefont{Linke}},
  \bibinfo{author}{\bibfnamefont{P.}~\bibnamefont{H\"{a}nggi}},
  \bibinfo{author}{\bibfnamefont{J.}~\bibnamefont{Rubi}}, \bibnamefont{and}
  \bibinfo{author}{\bibfnamefont{A.}~\bibnamefont{P\'{e}rez-Madrid}},
  \bibinfo{journal}{Phys. Rev. E} \textbf{\bibinfo{volume}{65}},
  \bibinfo{pages}{031104} (\bibinfo{year}{2002}).

\bibitem[{\citenamefont{Marchesoni}(1997)}]{Comment}
\bibinfo{author}{\bibfnamefont{F.}~\bibnamefont{Marchesoni}},
  \bibinfo{journal}{Phys. Lett. A} \textbf{\bibinfo{volume}{231}},
  \bibinfo{pages}{61} (\bibinfo{year}{1997}).

\bibitem[{\citenamefont{Borromeo and
  Marchesoni}(2000{\natexlab{a}})}]{ForwardBack}
\bibinfo{author}{\bibfnamefont{M.}~\bibnamefont{Borromeo}} \bibnamefont{and}
  \bibinfo{author}{\bibfnamefont{F.}~\bibnamefont{Marchesoni}},
  \bibinfo{journal}{Phys. Rev. Lett.} \textbf{\bibinfo{volume}{84}},
  \bibinfo{pages}{203} (\bibinfo{year}{2000}{\natexlab{a}}).

\bibitem[{\citenamefont{Machura et~al.}(2006)\citenamefont{Machura, Kostur,
  Talkner, Luczka, and H\"{a}nggi}}]{QDiff}
\bibinfo{author}{\bibfnamefont{L.}~\bibnamefont{Machura}},
  \bibinfo{author}{\bibfnamefont{M.}~\bibnamefont{Kostur}},
  \bibinfo{author}{\bibfnamefont{P.}~\bibnamefont{Talkner}},
  \bibinfo{author}{\bibfnamefont{J.}~\bibnamefont{Luczka}}, \bibnamefont{and}
  \bibinfo{author}{\bibfnamefont{P.}~\bibnamefont{H\"{a}nggi}},
  \bibinfo{journal}{Phys. Rev. E} \textbf{\bibinfo{volume}{73}},
  \bibinfo{pages}{031105} (\bibinfo{year}{2006}).

\bibitem[{\citenamefont{Miret-Artes and Pollak}(2005)}]{ExpActDyn}
\bibinfo{author}{\bibfnamefont{S.}~\bibnamefont{Miret-Artes}} \bibnamefont{and}
  \bibinfo{author}{\bibfnamefont{E.}~\bibnamefont{Pollak}},
  \bibinfo{journal}{J.Phys. Cond. Mat.} \textbf{\bibinfo{volume}{17}},
  \bibinfo{pages}{S4133} (\bibinfo{year}{2005}).

\bibitem[{\citenamefont{Lindenberg et~al.}(2005)\citenamefont{Lindenberg,
  Lacasta, and et~al}}]{NJP}
\bibinfo{author}{\bibfnamefont{K.}~\bibnamefont{Lindenberg}},
  \bibinfo{author}{\bibfnamefont{A.}~\bibnamefont{Lacasta}}, \bibnamefont{and}
  \bibinfo{author}{\bibfnamefont{J.~S.} \bibnamefont{et~al}},
  \bibinfo{journal}{New J.Phys.} \textbf{\bibinfo{volume}{7}},
  \bibinfo{pages}{29} (\bibinfo{year}{2005}).

\bibitem[{\citenamefont{Borromeo and Marchesoni}(2005)}]{NoiseMix}
\bibinfo{author}{\bibfnamefont{M.}~\bibnamefont{Borromeo}} \bibnamefont{and}
  \bibinfo{author}{\bibfnamefont{F.}~\bibnamefont{Marchesoni}},
  \bibinfo{journal}{Chaos} \textbf{\bibinfo{volume}{15}},
  \bibinfo{pages}{026110} (\bibinfo{year}{2005}).

\bibitem[{\citenamefont{Borromeo and Marchesoni}(2000{\natexlab{b}})}]{ACdrive}
\bibinfo{author}{\bibfnamefont{M.}~\bibnamefont{Borromeo}} \bibnamefont{and}
  \bibinfo{author}{\bibfnamefont{F.}~\bibnamefont{Marchesoni}},
  \bibinfo{journal}{Surf. Sci. Lett.} \textbf{\bibinfo{volume}{465}},
  \bibinfo{pages}{L771} (\bibinfo{year}{2000}{\natexlab{b}}).

\bibitem[{\citenamefont{Digal et~al.}(2000)\citenamefont{Digal, Ray, Sengupta,
  and Srivastava}}]{srivastava}
\bibinfo{author}{\bibfnamefont{S.}~\bibnamefont{Digal}},
  \bibinfo{author}{\bibfnamefont{R.}~\bibnamefont{Ray}},
  \bibinfo{author}{\bibfnamefont{S.}~\bibnamefont{Sengupta}}, \bibnamefont{and}
  \bibinfo{author}{\bibfnamefont{A.~M.} \bibnamefont{Srivastava}},
  \bibinfo{journal}{Phys. Rev. Lett.} \textbf{\bibinfo{volume}{84}},
  \bibinfo{pages}{826} (\bibinfo{year}{2000}).

\bibitem[{\citenamefont{Ferrando et~al.}(1995)\citenamefont{Ferrando,
  Spadacini, and Tommei}}]{UnbiasedSin}
\bibinfo{author}{\bibfnamefont{R.}~\bibnamefont{Ferrando}},
  \bibinfo{author}{\bibfnamefont{R.}~\bibnamefont{Spadacini}},
  \bibnamefont{and} \bibinfo{author}{\bibfnamefont{G.}~\bibnamefont{Tommei}},
  \bibinfo{journal}{Phys. Rev. E} \textbf{\bibinfo{volume}{51}},
  \bibinfo{pages}{126} (\bibinfo{year}{1995}).

\bibitem[{\citenamefont{H\"{a}nggi et~al.}(1990)\citenamefont{H\"{a}nggi,
  Talkner, and Borkovec}}]{RateTheory}
\bibinfo{author}{\bibfnamefont{P.}~\bibnamefont{H\"{a}nggi}},
  \bibinfo{author}{\bibfnamefont{P.}~\bibnamefont{Talkner}}, \bibnamefont{and}
  \bibinfo{author}{\bibfnamefont{M.}~\bibnamefont{Borkovec}},
  \bibinfo{journal}{Rev. Mod. Phys.} \textbf{\bibinfo{volume}{62}},
  \bibinfo{pages}{251} (\bibinfo{year}{1990}).

\bibitem[{\citenamefont{Reimann et~al.}(2001)\citenamefont{Reimann, den Broeck,
  Linke, H\"{a}nggi, Rubi, and P\'{e}rez-Madrid}}]{ODMobility}
\bibinfo{author}{\bibfnamefont{P.}~\bibnamefont{Reimann}},
  \bibinfo{author}{\bibfnamefont{C.~V.} \bibnamefont{den Broeck}},
  \bibinfo{author}{\bibfnamefont{H.}~\bibnamefont{Linke}},
  \bibinfo{author}{\bibfnamefont{P.}~\bibnamefont{H\"{a}nggi}},
  \bibinfo{author}{\bibfnamefont{J.}~\bibnamefont{Rubi}}, \bibnamefont{and}
  \bibinfo{author}{\bibfnamefont{A.}~\bibnamefont{P\'{e}rez-Madrid}},
  \bibinfo{journal}{Phys. Rev. Lett.} \textbf{\bibinfo{volume}{87}},
  \bibinfo{pages}{010602} (\bibinfo{year}{2001}).

\bibitem[{\citenamefont{Jung and H\"{a}nggi}(1989)}]{FTHarmonics}
\bibinfo{author}{\bibfnamefont{P.}~\bibnamefont{Jung}} \bibnamefont{and}
  \bibinfo{author}{\bibfnamefont{P.}~\bibnamefont{H\"{a}nggi}},
  \bibinfo{journal}{Europhys. Lett.} \textbf{\bibinfo{volume}{8}},
  \bibinfo{pages}{505} (\bibinfo{year}{1989}).

\bibitem[{\citenamefont{Siddiqi et~al.}(2004)\citenamefont{Siddiqi, Vijay,
  Pierre, Wilson, Metcalfe, Rigetti, Frunzio, and Devoret}}]{BifAmp}
\bibinfo{author}{\bibfnamefont{I.}~\bibnamefont{Siddiqi}},
  \bibinfo{author}{\bibfnamefont{R.}~\bibnamefont{Vijay}},
  \bibinfo{author}{\bibfnamefont{F.}~\bibnamefont{Pierre}},
  \bibinfo{author}{\bibfnamefont{C.}~\bibnamefont{Wilson}},
  \bibinfo{author}{\bibfnamefont{M.}~\bibnamefont{Metcalfe}},
  \bibinfo{author}{\bibfnamefont{C.}~\bibnamefont{Rigetti}},
  \bibinfo{author}{\bibfnamefont{L.}~\bibnamefont{Frunzio}}, \bibnamefont{and}
  \bibinfo{author}{\bibfnamefont{M.}~\bibnamefont{Devoret}},
  \bibinfo{journal}{Phys. Rev. Lett.} \textbf{\bibinfo{volume}{93}},
  \bibinfo{pages}{207002} (\bibinfo{year}{2004}).

\bibitem[{\citenamefont{Sias et~al.}(2007)\citenamefont{Sias, Zenesini,
  Lignier, Wimberger, Ciampini, Morsch, and Arimondo}}]{RET}
\bibinfo{author}{\bibfnamefont{C.}~\bibnamefont{Sias}},
  \bibinfo{author}{\bibfnamefont{A.}~\bibnamefont{Zenesini}},
  \bibinfo{author}{\bibfnamefont{H.}~\bibnamefont{Lignier}},
  \bibinfo{author}{\bibfnamefont{S.}~\bibnamefont{Wimberger}},
  \bibinfo{author}{\bibfnamefont{D.}~\bibnamefont{Ciampini}},
  \bibinfo{author}{\bibfnamefont{O.}~\bibnamefont{Morsch}}, \bibnamefont{and}
  \bibinfo{author}{\bibfnamefont{E.}~\bibnamefont{Arimondo}},
  \bibinfo{journal}{Phys. Rev. Lett.} \textbf{\bibinfo{volume}{98}},
  \bibinfo{pages}{120403} (\bibinfo{year}{2007}).

\bibitem[{\citenamefont{Tatarkova et~al.}(2003)\citenamefont{Tatarkova,
  Sibbett, and Dholakia}}]{CircBrownMotion}
\bibinfo{author}{\bibfnamefont{S.}~\bibnamefont{Tatarkova}},
  \bibinfo{author}{\bibfnamefont{W.}~\bibnamefont{Sibbett}}, \bibnamefont{and}
  \bibinfo{author}{\bibfnamefont{K.}~\bibnamefont{Dholakia}},
  \bibinfo{journal}{Phys. Rev. Lett.} \textbf{\bibinfo{volume}{91}},
  \bibinfo{pages}{038101} (\bibinfo{year}{2003}).

\end{thebibliography}

\end{document}